\documentclass[a4, showpacs, notitlepage, nofootinbib]{elsarticle}
\usepackage{graphicx}
\usepackage[siunitx]{circuitikz}
\usepackage{amsmath}
\usepackage{filecontents, url}
\usepackage{physics}
\usepackage{braket}
\usepackage{subcaption}
\usepackage{lipsum}
\usepackage{natbib}
\usepackage{mathtools}
\usepackage{footnote}
\usepackage{textcomp}
\usepackage{eqnarray,amsmath}
\usepackage{geometry}

\begin{document}

\title{Non-stationary GARCH modelling for fitting higher order moments of financial series within moving time windows}
\author{Luke De Clerk}
\address{Department of Physics, Loughborough University, Leicestershire, LE11 3TU, United Kingdom}
\ead{l.de-clerk@lboro.ac.uk}
\author{Sergey Savel'ev}
\address{Department of Physics, Loughborough University, Leicestershire, LE11 3TU, United Kingdom}
\ead{s.saveliev@lboro.ac.uk}
\date{\today}

\begin{abstract}
Here, we have analysed a GARCH(1,1) model with the aim to fit higher order moments for different companies' stock prices. When we assume a gaussian conditional distribution, we fail to capture any empirical data when fitting the first three even moments of financial time series. We show instead that a double gaussian conditional probability distribution better captures the higher order moments of the data. To demonstrate this point, we construct regions (phase diagrams), in the fourth and sixth order standardised moment space, where a GARCH(1,1) model can be used to fit these moments and compare them with the corresponding moments from empirical data for different sectors of the economy. We found that the ability of the GARCH model with a double gaussian conditional distribution to fit higher order moments is dictated by the time window our data spans. We can only fit data collected within specific time window lengths and only with certain parameters of the conditional double gaussian distribution. In order to incorporate the non-stationarity of financial series, we assume that the parameters of the GARCH model have time dependence.
\end{abstract}

\begin{keyword}
GARCH, Phase Diagrams, Double Gaussian, Empirical Data
\end{keyword}

\maketitle
\textit{JEL Classification: C10, C40, C50, C80}
\section{Introduction}
Modelling of financial time series is a very extensive area of research. One notable breakthrough in financial modelling is the discovery of the heteroskedasticity and conditional nature of volatility, manifesting itself in a slow stochastic process in the dynamics of price variance, in addition to the fast fluctuating price process itself. This motivates the development of the Autoregressive Conditional Heteroskedasticity models (ARCH) by Engle, \cite{arch}, which was later generalised (GARCH) by Bollerslev, \cite{GARCH1}. The autoregressive processes allow a stochastic model to predict price change probability density for a given time series. The level of return at a certain instance is described by a probability distribution (usually gaussian) and the variance of the process. Where the variance of the process varies with time and is defined by both the variance and level of return at the previous time instance(s). However, GARCH is not limited to simply financial systems but to any system where this two scale stochastic process is seen, for instance the study by Kumar et al. on atmospheric cycles in \cite{ozone} or the study on pathogen growth by Ali in \cite{pathogen_garch}.

Extensive research has been undertaken to adapt the original Bollerslev GARCH model to fit empirical observations of time series, \cite{engle_ap, gjr_garch, garch_optionpricing, garch_alternatives, nelson_garch, realised_1, realised_4}. In the majority of modelling, one assumes a stationarity of time series, however, this is not necessarily the case and as such the fractionally integrated GARCH (FIGARCH) models were proposed, \cite{figarch, figarch1, figarch2}. Nevertheless, whilst these modifications of GARCH increase the accuracy of forecasting volatility, there is an increase in the complexity of the models and in the ambiguity of estimating model parameters. For example, fitting higher order statistical moments of financial series is an attractive approach for estimating model parameters. The original GARCH model allows us to obtain analytical relations between the statistical moments and the GARCH model parameters. In contrast, the modifications of the GARCH model lead to an increase in the complexity of the expressions for the higher order moments making the evaluation of the model parameters impossible. Therefore, we wish to seek how effective the original GARCH model is at fitting higher order moments of empirical financial data series for different sectors of the economy. The focus of the paper upon higher order moments is primarily due to the higher order moments' ability to capture general aspects of a given distribution. It is well documented, \cite{4}, that higher order moments are able to quantitatively represent the number of events that differ largely from the mean value of the process. In essence, they provide a different way to capture and effectively describe the dynamics of the system.

The question is, therefore, how to fit prices for a given time series. We can set a task to fit a certain set of statistical moments, for example $\langle x^2\rangle$, $\langle x^4\rangle$ and $\langle x^6\rangle$ by the three parameters of the GARCH(1,1) model. Alternatively, we can choose another set $\langle x^2\rangle$, $\langle x^4\rangle$ and $\langle x^8\rangle$. Obviously, we will have different parameter values for the two cases. Moreover, we can ask if we can or cannot fit three empirically estimated moments of a chosen stock price series by three GARCH(1,1) parameters. The region where the parameters of the GARCH model can fit empirical moments shall be referred to throughout as the GARCH existence region or the `GARCHable' region. 

If we evaluate the time series and concludes that GARCH(1,1) parameters cannot fit empirical moments, then we can judge that the time series might no longer be purely stationary or a significant modification of the GARCH(1,1) model is needed. However, most time series are not purely stationary. As in the case of stocks and shares, the global economic climate and external factors are major stressors when determining the price of a given stock. Such a complex dependence of factors leads to a very fluid economic environment. Higher order moment analysis can determine the behaviour of the time series in response to this economic environment. In doing so, we can uncover a signal to these economic periods. However, economic factors affect the individual time series on different time scales. As such, the time window we analyse for the financial series will show different signatures for these different time scales as well as economic cycles and general tendencies. Therefore, it is plausible that truncating the time series into different time windows, we will gain different sets of model parameters for each time interval. The information about the changing behaviour of time series can manifest itself in a variation of the GARCH parameters and can identify changing economic factors and trends, including crisis periods, \cite{crash}. So, the final motivation for our work is to allow an individual to evaluate how best to model a series to gain the most appropriate set of parameters, if possible, of a GARCH(1,1) model for their particular needs.

The estimation data in this paper is for the period of 6th October 2000 to 6th October 2018, in most cases we use a subset of this data set. For example, this can be divided into a pre-crisis, post-crisis and crisis period. This division is extremely valuable in deducing the statistical features that are inherent to an economic crisis. This will be reflected in the results we gain from evaluating certain order moments in the years from 2000 to 2018.

The paper is organised as follows: in section \ref{raw}, we initially analyse the sixth order moment for several companies and discuss the economic environments. In section \ref{stoch}, we discuss the methods we will be using and how we have created the phase diagrams in higher order moment space where GARCH can describe empirical data. Section \ref{simu} presents our findings on a GARCH model with a gaussian conditional probability distribution (we will proceed calling these GARCH-normal models) for empirical time series fitting, whilst also showing the failure of GARCH-normal models to describe higher order moments of financial time series. In section \ref{double}, we discuss a GARCH model with a double gaussian conditional probability distribution (GARCH-double-normal models) to account for this shortfall. We also show how with the assumption of time dependent parameters, the data we analyse can be described by non-stationary GARCH-double-normal models. Section \ref{conc} concludes our paper.

\section{Raw Data Analysis}
\label{raw}
In order to determine the behaviour of the moments of financial time series, we first highlight the time dependence of the sixth order moment for several companies and a government bond (gilt) through the financial crisis of 2008. To do this, we use the daily closing price of each trading day over yearly periods for 8 years, 2002-2010. We then use the following equation to calculate the nth order moment: 

\begin{equation}
\langle x^n \rangle(t)=\frac{1}{N}\sum_{t-\frac{N}{2}\delta t<\tau<t+\frac{N}{2}\delta t} x^n(\tau)
\end{equation}
where we take $N$ to be a year, so the period we average over is 252 days (due to trading exclusion dates) and each time step $\delta t$ a trading day. Here we define $x$ as the logarithm of price change: 

\begin{equation}
x(\tau) = \ln \left( \frac{y(\tau+\delta t)}{y(\tau)}\right)
\end{equation}
$y(\tau)$ is the closing price at day $\tau$. In this paper, we also consider the average over a longer time history for example when $M = 4536$, in essence, we consider an 18 year time horizon: 

\begin{equation}
\langle x^n\rangle= \frac{1}{M}\sum_{i=1}^{M} x^n(t_0+i\delta t)
\label{mom}
\end{equation}
where, $t_0$ corresponds to the 6th October 2000. We also evaluate empirical standardised moments for higher order moments:

\begin{equation}
\Gamma_{2m}=\frac{\langle x^{2m}\rangle}{\langle x^2\rangle^m}
\label{genkurt}
\end{equation}

\begin{figure}[h!]
\centering
\includegraphics[width=\textwidth]{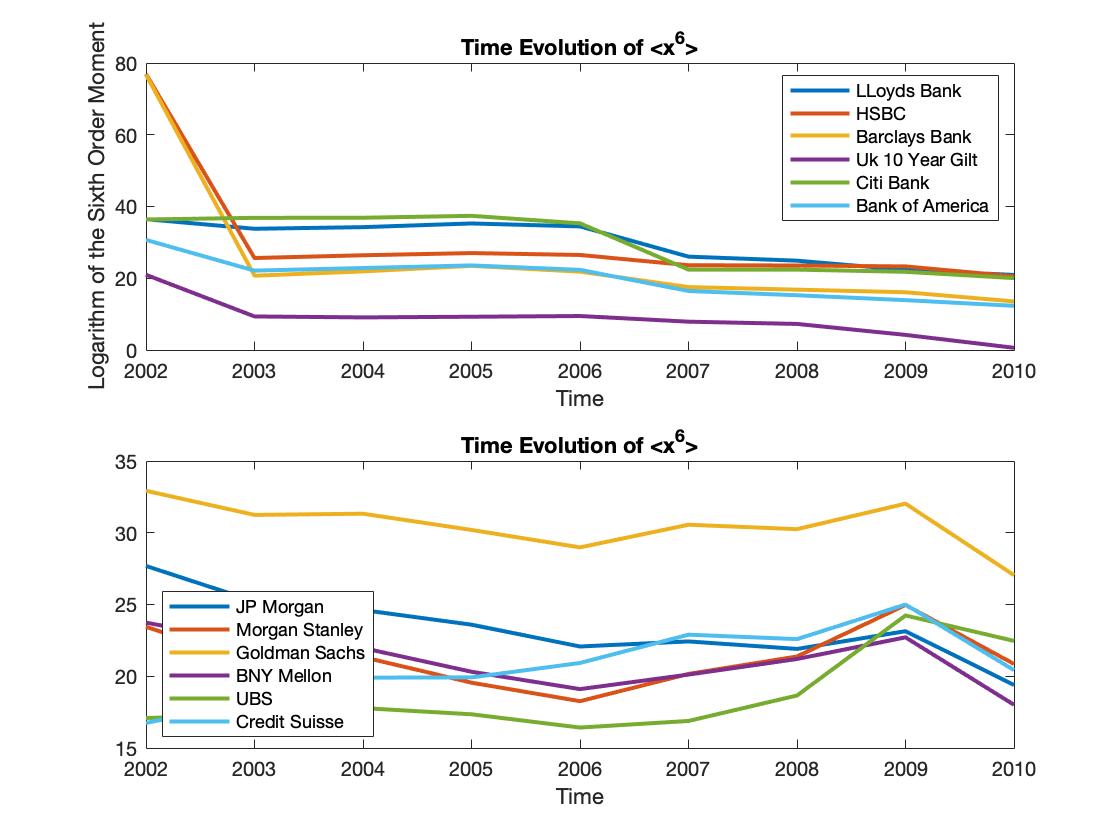}
\caption{Sixth order moment time evolution for several global companies, to highlight the behaviour of the sixth order moment in the financial crash of 2008. The top figure shows the evolution for primarily British companies and a gilt, whilst the bottom shows the same information for primarily American companies. The time period analysed for all companies spans 2002-2010.}
\label{raw1}
\end{figure}

From figure \ref{raw1}, we see that before the crisis (2002-2007) the sixth order statistical moments in general has a very flat dependence for the British companies. This is not mirrored in the American data as there is, in general, a negative rate of change for the sixth order moment.

What is evident from the yearly sixth moment values from 2002 to 2010, figure \ref{raw1}, is that within a financial crisis and subsequent economic collapse, the rate of change of the moments with respect to time increases in periods of financial worry. The nature of the rate of change, if it is positive or negative, can be attributed to the conditions the company encounters during the financial crisis. If we focus upon some specific companies we can highlight the circumstances that cause the nature of the rates of change. Firstly, during the 2007-2009 economic recession, Lloyds Bank was taken over by the UK Government. In figure \ref{raw1}, we can see that during the economic crisis, December 2007-June 2009, there is a negative rate of change of the sixth moment and as such the magnitude of the sixth moment is much lower after the economic recession, in 2009. This negative rate of change is seen in the UK 10 year Government bonds (gilts). Within the economic crisis, the Bank of England interest rates were decreased quite drastically and as such this increased the price of the bonds. This reduction in interest rates is a direct interference, as is the bailing out of a company, \cite{subprime}. We can see clearly that this nature of interference leads to a decrease in the sixth order statistical moment. Indicating the distribution describing the process has become narrower. Conversely, increasing the value of the sixth order moment gains a much broader distribution, highlighting increased uncertainty within the process. Therefore, looking at the behaviours of the sixth order moment throughout time horizons we can infer the behaviour of the distribution of the process.

Moreover, if we look at the non-British companies, the rate of change of the sixth order statistical moment is positive. The largest rate of change comes from Credit Suisse. Morgan Stanley, UBS, BNY Mellon, Goldman Sachs and HSBC also have positive rates of change of sixth order moment. These companies were not given such direct assistance and therefore the nature of these sixth order moments within this time period is indicative of an organic company within an economic crisis.

\section{Stochastic Model}
\label{stoch}
In this section, we focus upon a GARCH-normal(1,1) model. We can see from Bollerslev's work, \cite{GARCH1}, that for such a model $x_t$ is a random variable with zero mean and possesses the conditional variance, $\sigma_t^2$. We define $x_t\equiv \zeta_t\sigma_t$. Here, $\zeta_t$ is a random process with zero mean and variance equal to one. Depending on the system we wish to model, the variable $\zeta_t$ can be described by different probability distributions, see for example, \cite{garch_alternatives, continuous, levy_flight, cv, parameter}. However, as mentioned by Mantegna et al. in \cite{4}, it is often chosen to be gaussian. We shall first consider a gaussian random variable hence, we denote this model the GARCH-normal model. The GARCH(1,1) class of stochastic processes are defined via the relation:

\begin{equation}
\sigma_t^2=\alpha_0+\alpha_1x_{t-1}^2+\beta_1\sigma_{t-1}^2
\label{GARCH}
\end{equation}

If we knew the exact probability density, $p(x)$, of a process, we could write the definition of moments by, $E[x^{m}]=\int_{-\infty}^{\infty} P(x)x^m dx$. However, we do not know the analytical expression for the probability distribution of the GARCH process. To resolve this problem Bollerslev, \cite{GARCH1}, proposed recurrence relations for moments of the GARCH-normal(1,1) model:

\begin{equation}
E(x_t^{2m})=\frac{a_m\left[\sum_{n=0}^{m-1}a_n^{-1}(E(x_t^{2n}))\alpha_0^{m-n}{{m}\choose{m-n}} \mu(\alpha_1,\beta_1,n) \right]}{[1-\mu(\alpha_1,\beta_1,m)]}
\label{moments}
\end{equation}
where,

\begin{equation}
\mu(\alpha_1,\beta_1,m)=\sum_{j=0}^{m}{{m}\choose{j}}a_i\alpha_1^j\beta_1^{m-j} \; , \qquad a_j=\prod_{i=1}^{j}(2i-1)
\label{mu}
\end{equation}
Therefore, we can derive equations for the unconditional variance, the fourth order and sixth order standardised moments, \cite{4}:

\begin{equation}
\sigma^2 = \frac{\alpha_0}{1-\alpha_1-\beta_1}
\label{2}
\end{equation}

\begin{equation}
\Gamma_4 = \frac{E(x_t^4)}{E(x_t^2)^2} = 3 + \frac{6\alpha_1^2}{1-3\alpha_1^2-2\alpha_1 \beta_1-\beta_1^2}
\label{4}
\end{equation}

\begin{equation}
\Gamma_6=\frac{E(x_t^{6})}{(E(x_t^{2}))^3}=\frac{15(1-\alpha_1-\beta_1)^3(1+\frac{3(\alpha_1+\beta_1)}{1-\alpha_1-\beta_1}+\frac{3(1+\frac{2(\alpha_1+\beta_1)}{1-\alpha_1-\beta_1})(\beta_1^2+2\alpha_1\beta_1+3\alpha_1^2)}{1-3\alpha_1^2-2\alpha_1\beta_1-\beta_1^2})}{1-15\alpha_1^3-9\alpha_1^2\beta_1-3\alpha_1\beta_1^2-\beta_1^3}
\label{6}
\end{equation}

\begin{figure}
 \centering
\includegraphics[width=\linewidth]{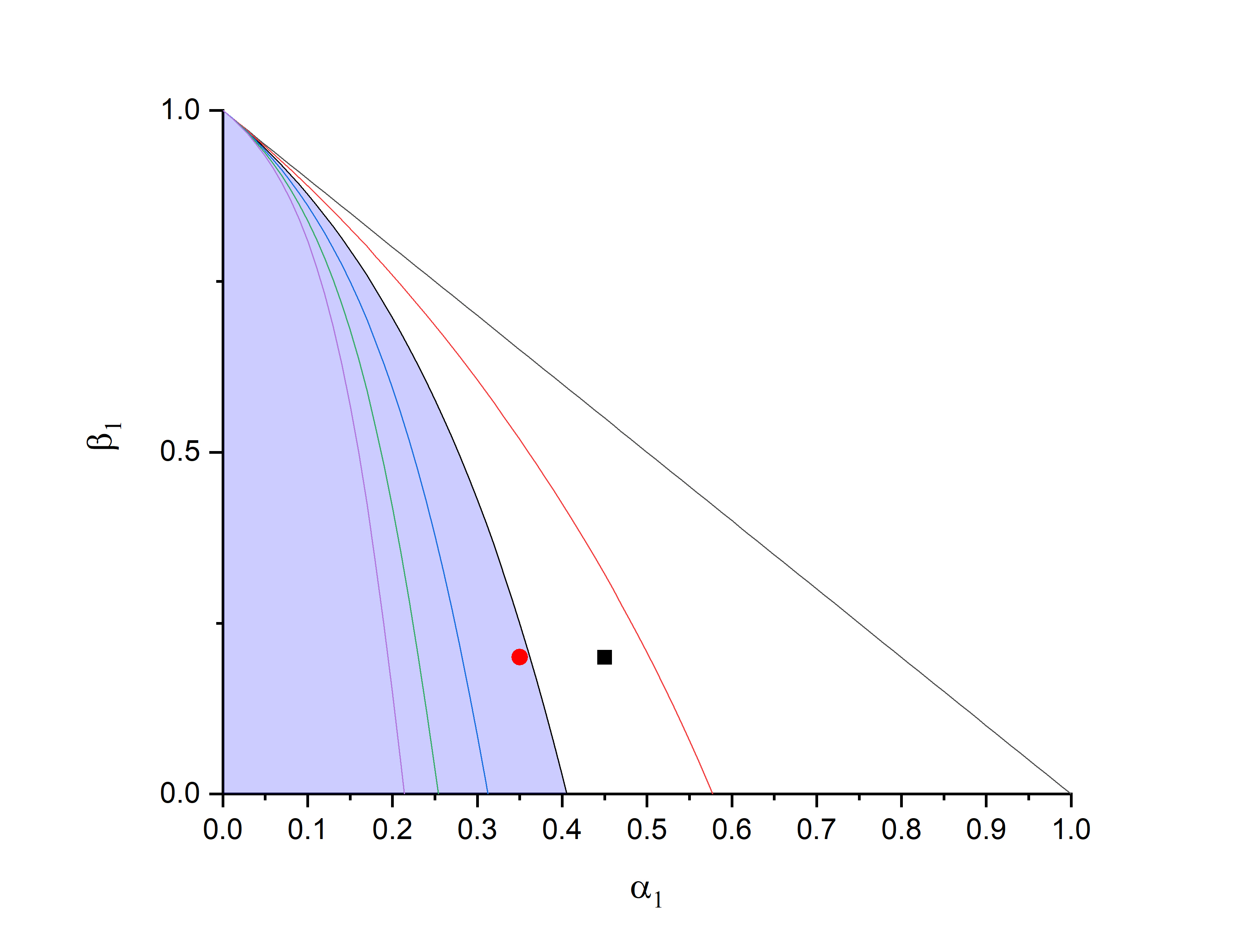}
\caption{The phase diagram, obtained in \cite{GARCH1}, showing the lines of divergence of moments in a GARCH-normal(1,1) model. The highlighted area shows the existence region for the sixth order moment. The red circle presents an example of $\alpha_1$, $\beta_1$ values that allow for the second, fourth and sixth order moments to exist, whilst the black square shows an example of $\alpha_1$, $\beta_1$ values that allow for only the second and fourth order moments to exist.}
\label{stability}
\end{figure}

The relation (\ref{moments}) and (\ref{mu}) are defined for all moments if we fix the three GARCH parameters, $\alpha_0$, $\alpha_1$ and $\beta_1$. For a moment to exist it is clear that we should have $\mu(\alpha_1, \beta_1,m) < 1$. Therefore, we can solve $\mu(\alpha_1, \beta_1, m)=1$ for $\beta_1=\beta_1^{(m)}(\alpha_1)$. In doing so, we can create figure \ref{stability}, \cite{GARCH1}. In this figure, we see the different curves of $\beta_1$=$\beta_1^{(m)}(\alpha_1)$ where $m$ takes the value; 2, 4, 6, 8, 10, and 12. When $\beta_1$$<$$\beta_1^{(m)}(\alpha_1)$ the corresponding moments $E[x_t^{2m}]$ and standardised moments $\Gamma_{2m}$ have finite values, whilst for $\beta_1$$>$$\beta_1^{(m)}(\alpha_1)$, these moments diverge. Since the particular line $\beta_1$$=$$\beta_1^{(m)}(\alpha_1)$  separates the region of parameters where the $2m$-th moment exists and where it does not, we can interpret this as a phase diagram in parameter space, \cite{GARCH1}. For the second, fourth and sixth order divergence lines we are able to gain analytical expressions for the divergence lines, see \ref{anal_divergence}. In figure \ref{stability}, we present a filled area that shows the region of existence of the sixth order moment. The red circle in this figure represents a set of parameter values that allow for the existence of the second, fourth and sixth moment but not the eighth or higher. While for the black square in figure \ref{stability} only the second and fourth order moments are finite. However, the inverse problem to estimate the three GARCH parameters, if three moments are known, is much more complicated and reduces to a set of transcendental equations which can be hard to solve.

\section{GARCH-Normal Models}
\label{simu}
\subsection{Company Trajectories}
Here, we will consider the situation of when we need to fit only the second and fourth order moments, or equivalently, fitting the unconditional variance $\langle x^2\rangle$ and fourth order standardised moment $\Gamma_4$. Since the GARCH-normal(1,1) model has three parameters we can conclude that we can express two GARCH parameters, for instance, $\alpha_1$ and $\beta_1$, as a function of the third parameters $\alpha_0$. To do so, we use equations (\ref{2}) and (\ref{4}) to fit the empirical values of variance, $\sigma_{emp}^2$, and the fourth order standardised moment, $\Gamma_{4, emp}$, for a certain company, such that $\sigma^2(\alpha_0, \alpha_1, \beta_1)=\sigma_{emp}^2$ and $\Gamma_4(\alpha_1, \beta_1)=\Gamma_{4, emp}$. In doing so, we derive:

\begin{equation}
\alpha_1=\sqrt{\frac{\frac{2\alpha_0}{\sigma^2_{emp}}-\frac{\alpha_0^2}{(\sigma^2_{emp})^2}}{\frac{6}{\Gamma_{4, emp}-3}+2}}
\label{12}
\end{equation}

\begin{equation}
\beta_1=1-\frac{\alpha_0}{\sigma^2_{emp}}-\sqrt{\frac{\frac{2\alpha_0}{\sigma^2_{emp}}-\frac{\alpha_0^2}{(\sigma^2_{emp})^2}}{\frac{6}{\Gamma_{4, emp}-3}+2}}
\label{11}
\end{equation}

It is clear from these equations that for any value of $\Gamma_{4, emp} > 3$ and $\sigma_{emp}^2 > 0$, we can find a family of one-parametric GARCH models. So, we obtain the parametric curves; $(\alpha_1(\alpha_0), \beta_1(\alpha_0))$, in $(\alpha_1$, $\beta_1)$ space. Such curves represent the `company trajectories' with already fixed (empirical) variance, $\sigma^2_{emp}$ and empirical fourth order standardised moments, $\Gamma_{4, emp}$.

\begin{figure*}
        \centering
        \begin{subfigure}[b]{0.48\textwidth}
            \centering
            \includegraphics[width=\textwidth]{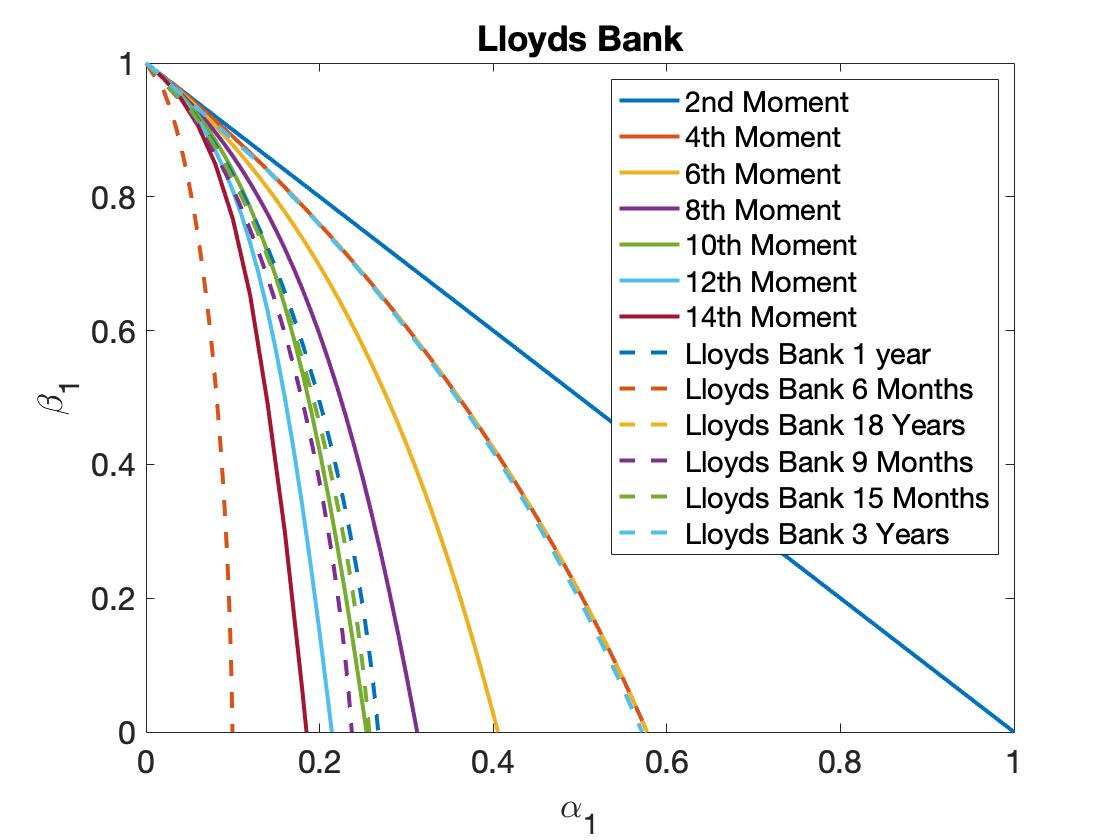}
            \caption[Network2]%
            {{\small}}    
            \label{lloydsyear}
        \end{subfigure}
        \hfill
        \begin{subfigure}[b]{0.48\textwidth}  
            \centering 
            \includegraphics[width=\textwidth]{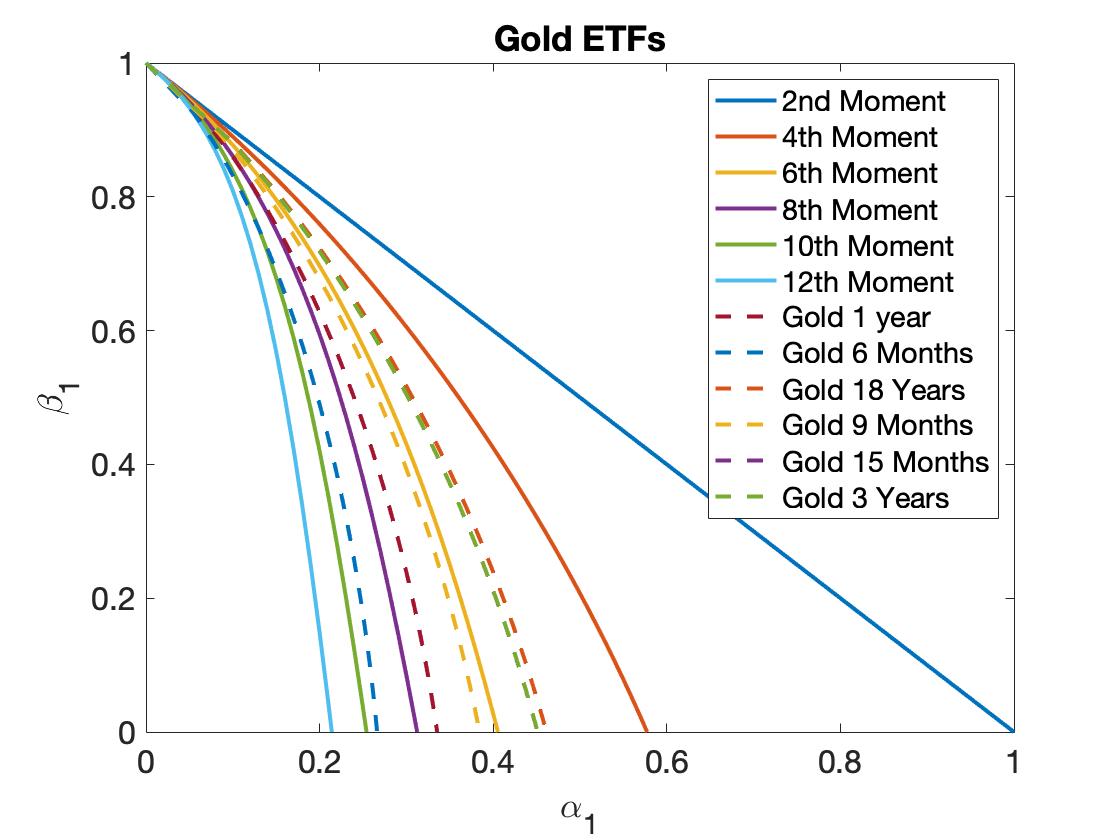}
            \caption[]%
            {{\small}}    
            \label{goldyear}
        \end{subfigure}
        \vskip\baselineskip
        \begin{subfigure}[b]{0.48\textwidth}   
            \centering 
            \includegraphics[width=\textwidth]{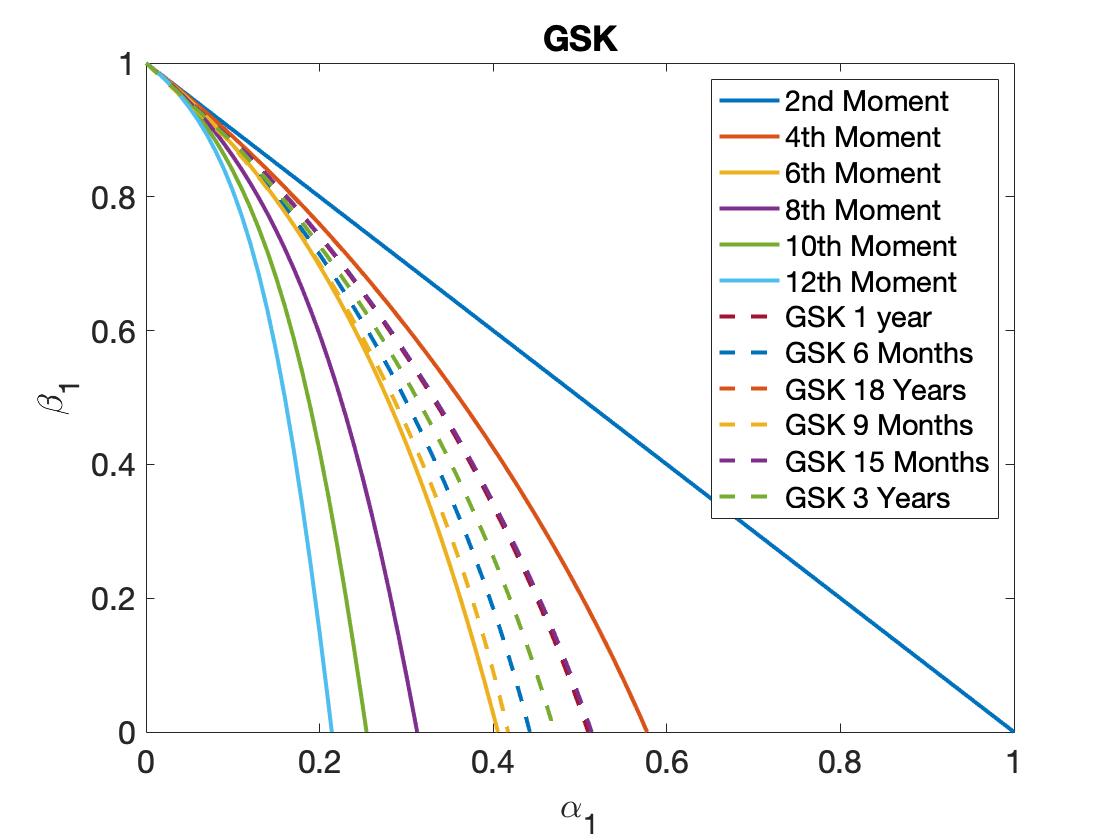}
            \caption[]%
            {{\small}}    
            \label{gskyear}
        \end{subfigure}
        \hfill
        \begin{subfigure}[b]{0.48\textwidth}   
            \centering 
            \includegraphics[width=\textwidth]{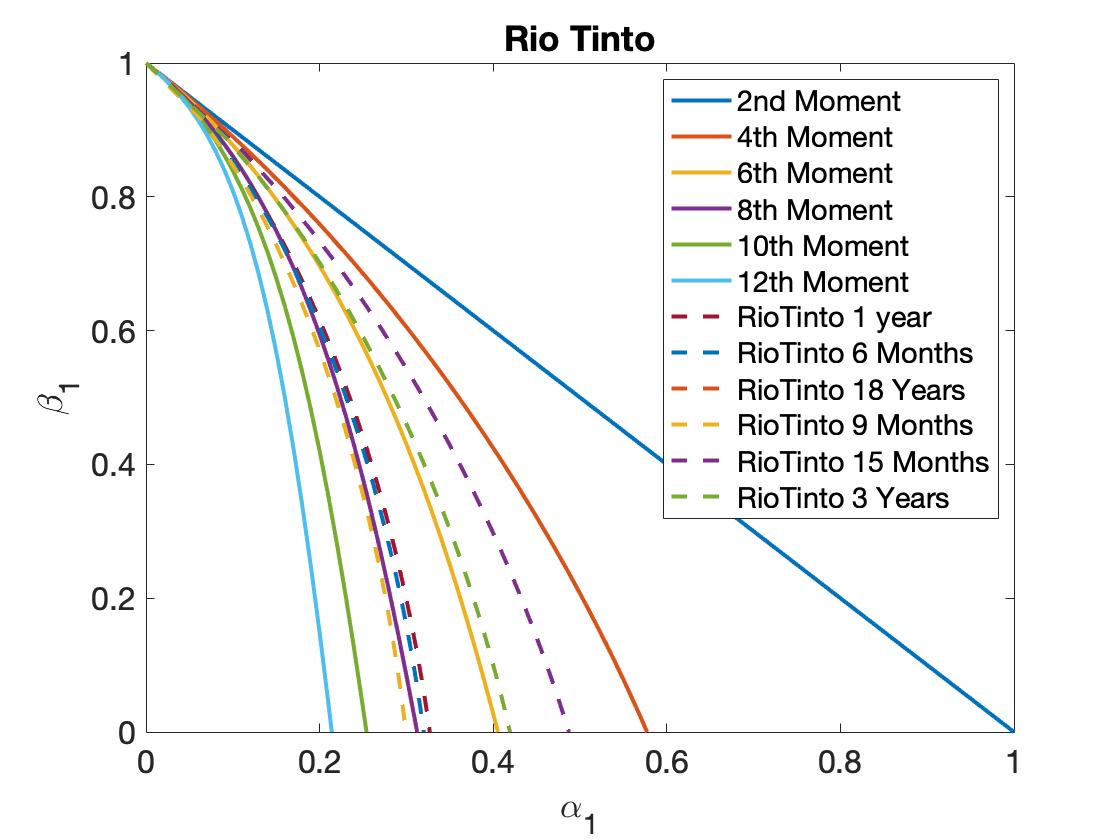}
            \caption[]%
            {{\small}}    
            \label{rtyear}
        \end{subfigure}
        \caption[]
        {\small The stability phase diagram for the GARCH-normal(1,1) moments with an overlap of several company trajectories. Panel (a) shows the trajectory for Lloyds Bank, where the shortest time window allows up to the twelfth order moment to exist. Panel (b) shows the trajectory of Gold ETFs (Exchange Traded Funds), where the shortest time window allows to the eighth moment to exist. Panel (c) shows the same for GSK, which allows up to the fourth moment to exist and lastly, panel (d) shows the same for Rio Tinto, which allows up to the eighth order moment to exist.} 
        \label{stability1}
 \end{figure*}

In figure \ref{stability1}, we see an extension of figure \ref{stability} for a banking stock, a commodity, a pharmaceutical and a mining company, respectively. The dotted lines represent the parameters of the GARCH-normal model for the given company's trajectory. They allow us to see the `stability' of the time series, in essence, which statistical moments can exist for the GARCH description of the empirical data of a certain company. It is evident, for the longest time period (18 years) that apart from the gold ETFs (Exchange Traded Funds), trajectories of all other companies lie above the divergence line of the sixth order moment. Implying, the values of the second and fourth order empirical statistical moments do not allow for any higher order moments to be fitted via a GARCH-normal model.

If we decrease the time window of data collection, for example a year, 6th October 2017 to 6th October 2018, or even six months, 6th April 2018 to 6th October 2018, then we can see the migration of the company trajectory to deeper inside the stability region in the $(\alpha_1$, $\beta_1$) plane, where higher moments are finite (see figures \ref{lloydsyear}, \ref{goldyear}, \ref{gskyear}, \ref{rtyear}). We have also examined the time windows of nine months, fifteen months and three years.  In these figures (figures \ref{lloydsyear}, \ref{goldyear}, \ref{gskyear}, \ref{rtyear}), it is clear that the Lloyds Bank 6 month time series allows the largest number of higher order moments to exist for its description within the corresponding GARCH-normal(1,1) model. Apart from Lloyds Bank all other time series present only with the fourth or eighth moment as the highest finite moment for the GARCH-normal model along the corresponding company trajectories. However, in general, it is clear to see that the shorter a time series we take, the more moments exist for a GARCH-normal(1,1) model.

As we traverse a company's trajectories in ($\alpha_1$, $\beta_1$) space we can work out the value of the sixth order standardised moment generated from the GARCH-normal(1,1) model for these specific $\alpha_1$ and $\beta_1$ values. In table \ref{g6_min}, we see the minimum and maximum of $\Gamma_6$ generated. We can see in general $\Gamma_6$ does not vary significantly along the company trajectory, resulting in a problem to fit diverse values of the empirical sixth order standardised moments.

\begin{table}[h!]
\centering
\begin{tabular}[c]{|p{3cm}|p{2cm}|p{2cm}|}
\hline
Company&Minimum&Maximum\\ \hline
Lloyds- 6 Months&33.3700&37.0000\\ \hline
GSK- 6 Months&60.3386&76.5162\\ \hline
Gold- 3 Year&90.3439&253.4463\\ \hline
Gold- 1 Year&79.3684&188.9430\\ \hline
RioTinto- 3 Year&36.1730&42.5383\\ \hline
RioTinto- 1 Year&34.1419&39.9972\\ \hline
RioTinto- 6 Months&30.9564&35.3291\\ \hline
\end{tabular}
\caption{The minimum and maximum values of $\Gamma_6$ along the company's trajectories in ($\alpha_1$, $\beta_1$) space.}
\label{g6_min}
\end{table}

\subsection{Methods of Parameter Fitting}
If we want to fit the second, fourth and sixth moments, the values of the parameters must be below the divergence curve; $\beta_1<\beta_1^{(6)}(\alpha_1)$, which does not cover all parameter space for the existence of the fourth ($\beta_1<\beta_1^{(4)}$) and the second ($\beta_1<\beta_1^{(2)}$) order moments. This can result in some values of the fourth and second moment, or fourth order standardised moment and the second order moment being unreachable for GARCH modelling, see \ref{divergence_proof}. 

Let us consider the algorithms we can use to fit empirical values of $\langle x^2 \rangle, \langle x^4 \rangle$ and $\langle x^6 \rangle$ which can be reformulated as: the variance, $\sigma^2_{emp}$, as well as the fourth and sixth order standardised moments, $\Gamma_{4, emp}$ and $\Gamma_{6, emp}$, respectively. In the first approach, we present $\alpha_1$ and $\beta_1$, as a function of $\alpha_0$, that is $\alpha_1(\alpha_0,\sigma_{emp}^2,\Gamma_{4,emp})$ and $\beta_1(\alpha_0,\sigma_{emp}^2,\Gamma_{4,emp})$, from equations (\ref{12}) and (\ref{11}), then numerically solve the equation:

\begin{equation}
\Gamma_6(\alpha_1(\alpha_0, \sigma^2_{emp}, \Gamma_{4, emp}), \beta_1(\alpha_0,\sigma^2_{emp},\Gamma_{4, emp}), \alpha_0)=\Gamma_{6, emp}
\end{equation}
to find the value of $\alpha_0$. This method is inspired by the trajectory analysis we use in the previous section. We search for $\alpha_0$ by traversing the trajectory and trying to fit the empirical sixth order standardised moment. However, if $\Gamma_6$ is lower than the minimum or larger than the maximum of possible $\Gamma_6$ stated in table \ref{g6_min}, this equation cannot be solved. Indicating that the GARCH-normal model with such a value of the empirical sixth order standardised moment does not exist.

In the second approach to fit empirical values of $\langle x^2\rangle$, $\Gamma_4$ and $\Gamma_6$, we first fit the empirical fourth and sixth order standardised moments using the fact that $\Gamma_4(\alpha_1, \beta_1)$ and $\Gamma_6(\alpha_1, \beta_1)$ do not depend on $\alpha_0$, see equations (\ref{4}) and (\ref{6}). Therefore, we can reduce the problem to two equations:

\begin{equation}
\begin{aligned}
&\Gamma_{4}(\alpha_1, \beta_1) = \Gamma_{4, emp}, \\
&\Gamma_6(\alpha_1, \beta_1) = \Gamma_{6, emp}
\label{eqt_array}
\end{aligned}
\end{equation}
allowing us to evaluate values of $\alpha_1, \beta_1$ and reserve $\alpha_0$ to the fitting of variance; $\alpha_0 = \sigma^2_{emp}(1-\alpha_1-\beta_1)$. The set of equations (\ref{eqt_array}) can be further reduced to one equation by eliminating $\beta_1$ using the first equation of the set namely:

\begin{equation}
\beta_1 = \sqrt{1-2\alpha_1^2-\frac{6\alpha_1^2}{\Gamma_{4, emp} - 3}} - \alpha_1
\label{b1g4}
\end{equation}
and substituting it to the second equation of (\ref{eqt_array}). This enables us to write the one-variable equation:

\begin{equation}
\Gamma_6(\beta_1(\alpha_1, \Gamma_{4, emp}), \alpha_1) = \Gamma_{6, emp}
\end{equation}
Note, we similarly can exclude $\alpha_1$, resulting in equations for $\beta_1$.

\subsection{Phase Diagram}
\begin{figure}
\centering
\includegraphics[width=\linewidth]{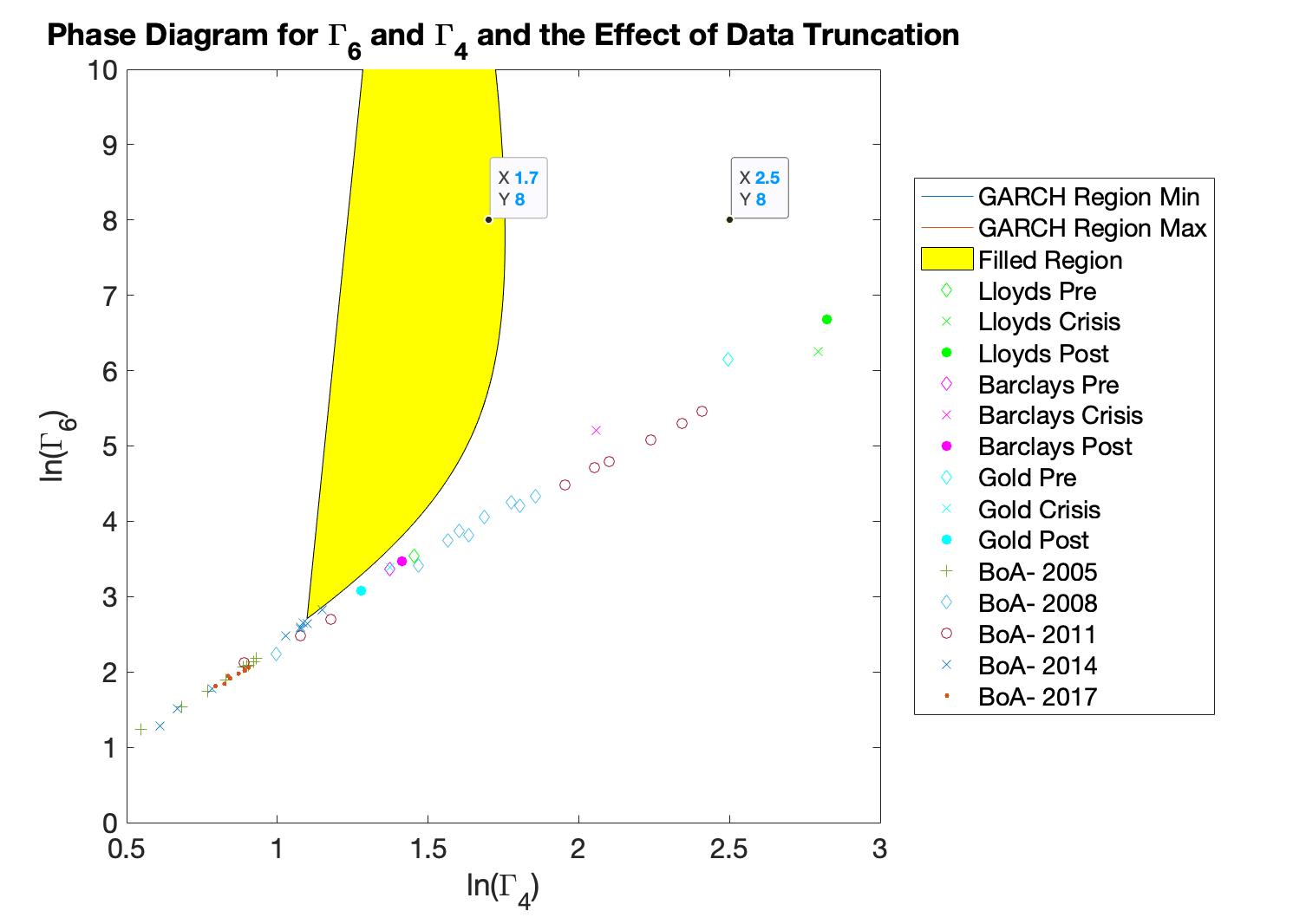} 
\caption{The phase diagram for ($\ln(\Gamma_4)$, $\ln(\Gamma_6)$) space. The filled yellow area shows the region in which the GARCH-normal model is able to fit the fourth and sixth order standardised moment, whilst the rest of the space is where the certain values of $\Gamma_4$ and $\Gamma_6$ cannot be fitted via a GARCH-normal(1,1) model. The two highlighted points $(1.7, 8)$ and $(2.5, 8)$ show the examples of values of $\Gamma_4$ and $\Gamma_6$ that can be fitted by a GARCH model and those that cannot, respectively. The other data points in the space, represent the empirical data for several companies, truncated to 10-90 day windows, incremented in 10 day periods. It is shown here that no empirical data can be fitted via a GARCH-normal(1,1) model.}
\label{boa_truncated}
\end{figure}

The equations (\ref{eqt_array}) can only be solved for some region, in standardised moment space, ($\Gamma_4$, $\Gamma_6$), shaded in yellow in figure \ref{boa_truncated}. This is the region of phase space where the respective values of the fourth and sixth order standardised moments can be fitted via a GARCH-normal model. For example, the first point ($1.7, 8$) is inside the `GARCHable' region. Whilst, the second point ($2.5, 8$), is outside of the `GARCHable' region, highlighting that these moment values cannot be fitted via a GARCH-normal model. Therefore, no solution is possible to the equations (\ref{eqt_array}).

\begin{figure}[h!]
\centering
\includegraphics[width=\textwidth]{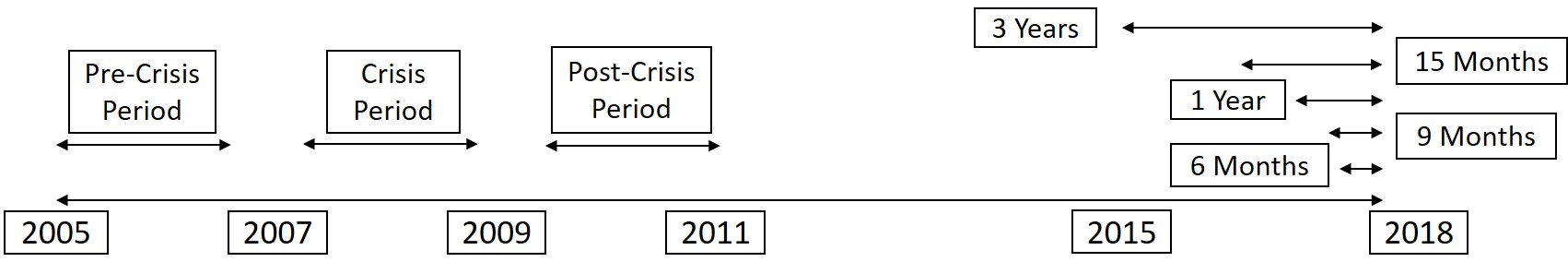}
\caption{The timeline of time series windows that we investigate within the paper. We investigate periods of economic turmoil; pre-crisis, crisis and post-crisis periods, as well as relatively stable periods. We also highlight here the truncation of 3 years, 15 months, 1 year, 9 months and 6 months.}
\label{timeline}
\end{figure}

To evaluate the appropriateness of a GARCH-normal(1,1) model for the fitting of higher order moments in stock market data, we shall be investigating time series for companies of different sectors of the economy by estimating their empirical values of the fourth and sixth order standardised moments and comparing to the GARCH-normal(1,1) phase region in ($\Gamma_4$, $\Gamma_6$) space, figure \ref{boa_truncated}. To see the effect of the length of the time window on the ability of the GARCH-normal(1,1) model to fit empirical moments, we divide data in to different economic periods. In figure \ref{timeline}, we show the different regions we wish to evaluate. We start with taking six month periods; 6th April to the 6th October in the years 2005, 2008, 2011, 2014 and 2017. We break these six month time periods in to 10, 20, 30, 40, 50, 60, 70, 80 and 90 day windows, an example of this can be seen for the several stocks in figure \ref{boa_truncated}, \cite{luke_emp}. We then overlap these data points on top of the `GARCHable' region detailed above.

We do not see the empirical data inside of the GARCH-normal(1,1) phase region for the time periods analysed. Therefore, we can say that a GARCH-normal(1,1) model is unable to fit higher order moments of the empirical time series we have studied.
 
\section{GARCH-Double-Normal Models}\label{double}
It is clear from this initial analysis of the GARCH-normal model that a gaussian conditional distribution does not allow a fitting for both the fourth and sixth standardised moments for stock data. This problem can potentially be resolved if we replace the conditional gaussian distribution with a distribution allowing to have a better flexibility in higher order GARCH standardised moments. Therefore, we seek a distribution with a larger fourth order standardised moment value in the region where $\Gamma_6$ exists. We introduce the double gaussian distribution \cite{double_gauss}, where $p(x)$:

\begin{equation}
p(x) = ae^{\frac{-x^2}{\sigma_1^2}} + be^{\frac{-x^2}{\sigma_2^2}}
\label{p_double}
\end{equation}
In addition to an obvious normalisation condition: 

\begin{equation}
a+b=1
\end{equation}
we also have constraints on the 2nd moment: 

\begin{equation}
E[x^2] = a\sigma_1^2 + b\sigma_2^2 = 1
\label{double_2}
\end{equation}
due to the requirement that the conditional distribution for a GARCH process should have variance equal to one. We can introduce two more variables (higher order moments; 4th and 6th moments of the conditional distribution) which fully define all parameters in equation (\ref{p_double}):

\begin{equation}
E[x^4] = a\sigma_1^4 + b\sigma_2^4 = \mu_4 = \frac{\eta_4}{3}
\label{double_4}
\end{equation}

\begin{equation}
E[x^6] = a\sigma_1^6 + b\sigma_2^6 = \mu_6 = \frac{\eta_6}{15}
\label{double_6}
\end{equation}

The parameterisation (\ref{double_4}) and (\ref{double_6}) of the double gaussian distribution allows us to generalise Bollerslev's equation (\ref{4}). The second order moment is not affected and is still determined by equation (\ref{2}), while the fourth and sixth order standardised moments for GARCH with double-gaussian distribution can be written as:
 
\begin{equation}
\Gamma_4 = \frac{\eta_4(1-\alpha_1-\beta_1)^2(1 + \frac{2(\alpha_1+\beta_1)}{1-\alpha_1-\beta_1})}{1-\eta_4\alpha_1^2-2\alpha_1\beta_1-\beta_1^2}
\label{general1}
\end{equation}
 
\begin{equation}
\Gamma_6 = \frac{\eta_6(1-\alpha_1-\beta_1)^3(1+\frac{3(\alpha_1+\beta_1)}{1-\alpha_1-\beta_1}+3(1+\frac{2(\alpha_1+\beta_1)(\eta_4\alpha_1^2+2\alpha_1\beta_1+\beta_1^2)}{1-\eta_4\alpha_1-2\alpha_1\beta_1-\beta_1^2}))}{1-\eta_6\alpha_1^3-3\eta_4\alpha_1^2\beta_1-3\alpha_1\beta_1^2-\beta_1^3}
\label{general}
\end{equation}

\begin{figure}[h!]
\centering
\includegraphics[width=\textwidth]{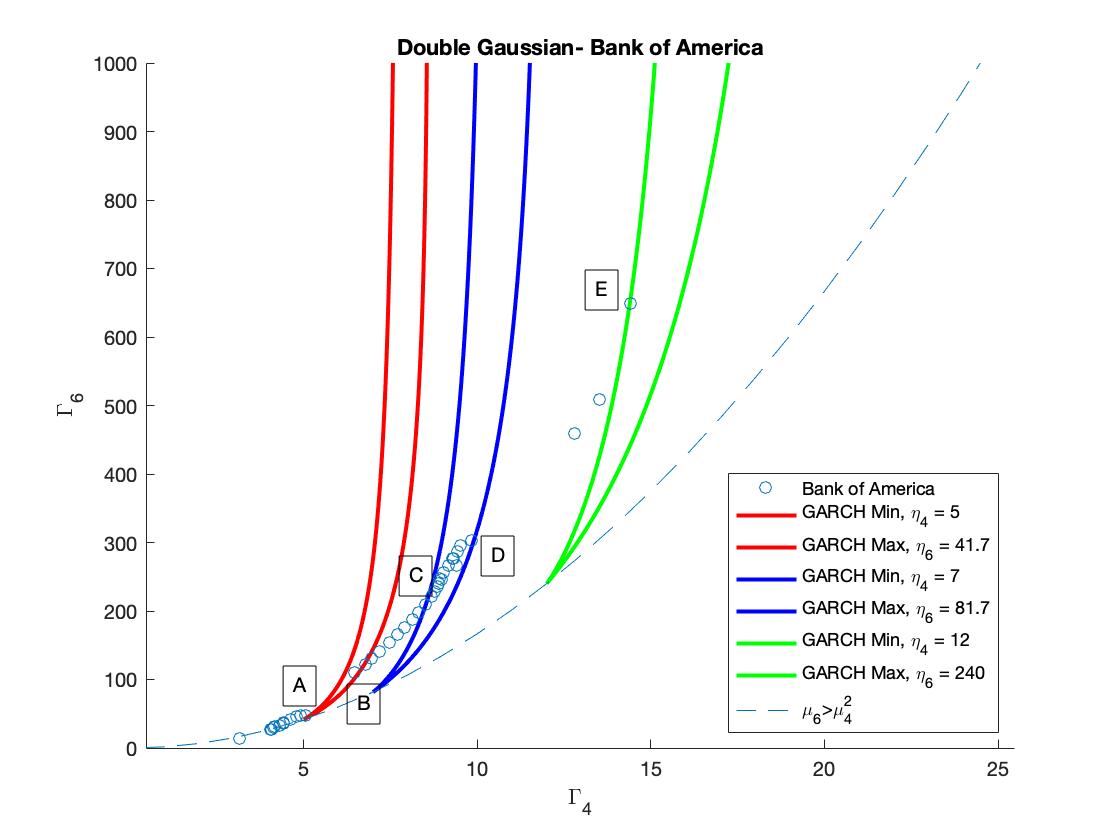}
\caption{The phase diagrams for ($\Gamma_4$, $\Gamma_6$) space for GARCH-double-normal(1,1) models corresponding to different parameters given in table \ref{phase2}. From equations (\ref{double_4}) and (\ref{double_6}), we are able to gain conditions that restrict the values of $\eta_4$ and $\eta_6$ to describe a double gaussian distribution. We then overlay the empirical data for Bank of America, truncated from $1\%$ to $100\%$ of the length of the time series, incremented in one percent steps, for the period of 6th October 2000 to 6th October 2018. To highlight the ability of the GARCH-double-normal model to fit higher order moments for specific lengths of time windows we present three regions for the space that allow the fitting of the fourth and sixth order standardised moments by a GARCH-double-normal model. Each has a different time window that it can fit, shown by the letter, associated with table \ref{phase2}.}
\label{phase_double}
\end{figure}

Using the methods described prior and based on the existence of solutions of the set of equations (\ref{eqt_array}), as well as, equations (\ref{general1}) and (\ref{general}), we create a family of phase diagrams parameterised by $\eta_4$ and $\eta_6$. To understand which empirical values are achievable using a GARCH-double-normal model, we need to understand restrictions for the whole family of phase diagrams. We see that these are bounded due to limitations for $\eta_4$ and $\eta_6$ obtained in \ref{double_gaussian} (conditions \ref{con1} and \ref{con2}). These limitations require all phase diagrams be started from points above or on the dashed line, figure \ref{phase_double}. As such only data above or on the dashed line can be described by a GARCH-double-normal model (which is the case for the empirical data collected for the securities we have considered here).

\subsection{Time Windows}
In figure \ref{phase_double}, we see three phase diagrams for three different double gaussian distributions. Parameters for these phase diagrams are given in table \ref{phase2}.

\begin{table}[h!]
\centering
\begin{tabular}[c]{|p{2cm}|p{2cm}|p{2cm}|p{2cm}|p{2cm}|}
\hline
Position&$\eta_4$&$\eta_6$&$t_{min}(Days)$&$t_{max}(Days)$\\ \hline
Leftmost (Red)&5&41.7&171 (A)&600 (B)\\ \hline
Centre (Blue)&7&81.7&943 (C)&1500 (D) \\ \hline
Rightmost (Green)&12&240&1586 (E)&1586 \\ \hline
\end{tabular}
\caption{Parameters of the conditional double gaussian distributions used to construct `GARCHable' regions in figure \ref{phase_double}. The table summarises the parameters of the distributions used to model the time windows (A) to (B), (C) to (D) and (E). These are the limits of the time windows in days that the particular instance of the double gaussian distribution can be used to fit the higher order moments of the empirical data of the Bank of America.}
\label{phase2}
\end{table}
Figure \ref{phase_double} demonstrates how altering the parameters $\eta_4$ and $\eta_6$ of the GARCH-double-normal model enables us to capture different time windows of the empirical data. The data used in figure \ref{phase_double} is for the Bank of America time series from 6th October 2000 to 6th October 2018. We truncate the time series in to different lengths. We start with $1\%$ of the overall length and increment by $1\%$ up to the whole length of the time series. In other words, the first, most left point, corresponds to 43 days of data, (from 06/10/2000 to 04/12/2000), the second point corresponds to moments obtained for 86 days of data (from 06/10/2000 to 31/01/2001) and so on.

For the leftmost phase diagram we use a double gaussian distribution with $\eta_4 = 5, \eta_6 = 41.7$. This allows us to fit $\sigma^2$, $\Gamma_4$ and $\Gamma_6$ for the time window of duration in the interval, $171 \leq t \leq 600$ days. When fitting higher order moments for longer time windows, we need to use double gaussian distributions with parameters summarised in table \ref{phase2}. It is not possible to gain a GARCH process with a double gaussian distribution to capture all of the empirical data's higher order standardised moments. We denote this behaviour as the local ability to model higher order moments of financial time series by the GARCH-double-normal model. Figure \ref{phase_double} only uses the data from the Bank of America, however, this behaviour is seen throughout the empirical data we have studied. In order to capture the empirical data we must first decide on the time window we wish to model and then ascertain a suitable distribution that will capture this window.

\subsection{Time Dependence of GARCH-double-normal Parameters}
\begin{figure}[h!]
\centering
\includegraphics[width=\textwidth]{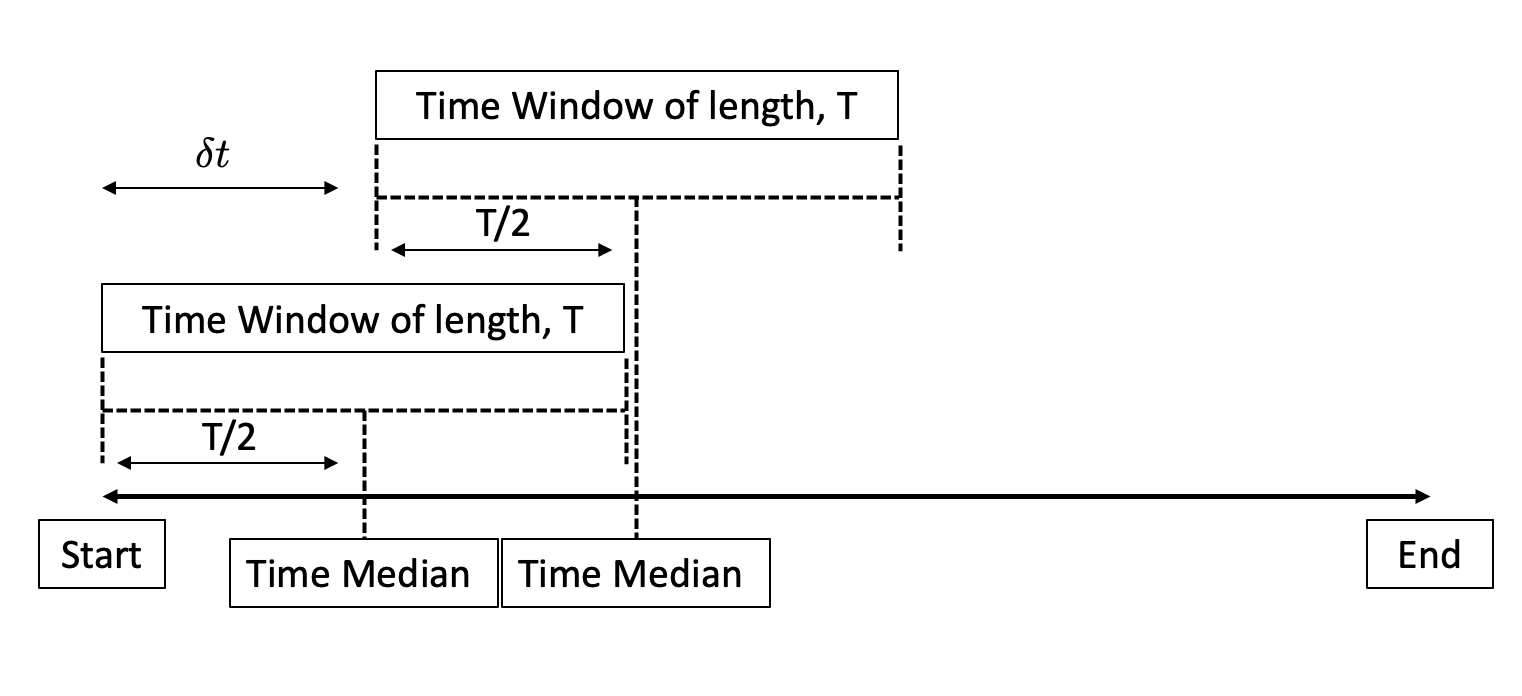}
\caption{Here, we show how we take a rolling window for a long time series (from the start to the end). We highlight two fixed time windows, of certain length, T, with time median that correspond to the middle of these windows. This time window then shifts in time, by time $\delta t$, taken in this paper to be a day.}
\label{timewindow}
\end{figure}

Once we have fixed the time window we wish to model, we can study what happens when the window with this fixed duration shifts in time. This can be done by attributing to the higher order moments a time moment, $t$, corresponding to the middle point (the median) of the time window we wish to model. This can be seen in figure \ref{timewindow}, where we detail the schematic of a fixed time window moving in time for a long time series.

If we fix the double gaussian distribution (in essence, select certain $\eta_4$ and $\eta_6$), we can gain the set of GARCH parameters, $\alpha_0$, $\alpha_1$ and $\beta_1$ that describes the particular time median. If we change the time window we look at by moving its time median, then the GARCH parameters $\alpha_0$, $\alpha_1$ and $\beta_1$ also change. Below, we observe that the GARCH parameters $\alpha_0(t)$, $\alpha_1(t)$ and $\beta_1(t)$ significantly vary with the moving time window, highlighting the non-stationarity of our modelling.

Given equations (\ref{2}) and (\ref{general1}), we are able to define trajectories in ($\alpha_1$, $\beta_1$) space for a fixed value of $\sigma^2_{emp}$ and $\Gamma_{4, emp}$. Unlike the GARCH-normal methods, we now have the trajectories as a function of $\eta_4$. These can be seen below:

\begin{equation}
\alpha_1 = \sqrt{\frac{(\frac{2\alpha_0}{\sigma_{emp}^2}-(\frac{\alpha_0}{\sigma^2_{emp}})^2)(\Gamma_{4, emp}-\eta_4)}{(\Gamma_{4, emp}-\eta_4)(\eta_4-1)-(\eta_4^2-\eta_4)}}
\label{a1d}
\end{equation}

\begin{equation}
\beta_1 = 1 - \frac{\alpha_0}{\sigma_{emp}^2} - \sqrt{\frac{(\frac{2\alpha_0}{\sigma_{emp}^2}-(\frac{\alpha_0}{\sigma^2_{emp}})^2)(\Gamma_{4, emp}-\eta_4)}{(\Gamma_{4, emp}-\eta_4)(\eta_4-1)-(\eta_4^2-\eta_4)}}
\label{b1d}
\end{equation}

\begin{figure}[h!]
\centering
\includegraphics[width=0.75\textwidth]{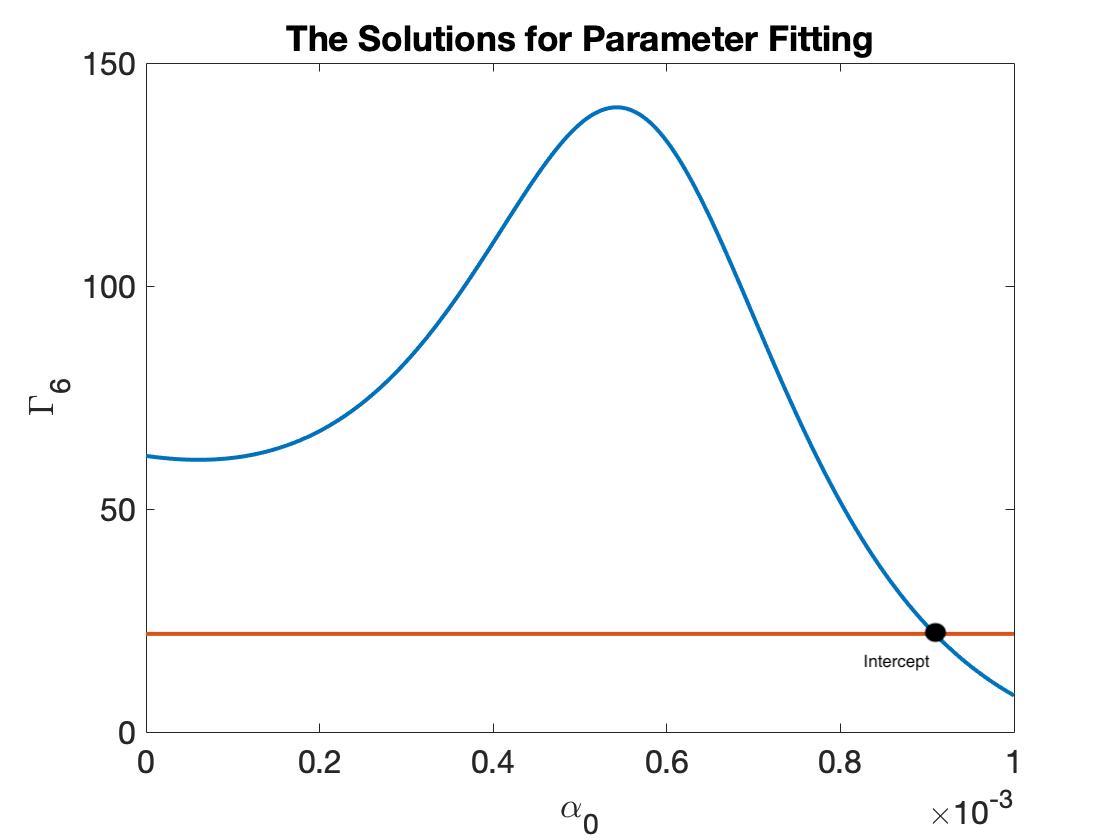}
\caption{$\Gamma_6$ as a function of the running parameter $\alpha_0$. Here, we show how we calculate the value of $\alpha_0$ for a particular time window. The orange line is the value of $\Gamma_6$ for the empirical time window we are modelling, whilst the blue line shows $\Gamma_6(\alpha_0)$, equation (\ref{general}) when $\alpha_1=\alpha_1(\eta_4, \alpha_0, \sigma^2_{emp}, \Gamma_{4, emp})$ (equation (\ref{a1d})) and $\beta_1=\beta_1(\eta_4, \alpha_0, \sigma^2_{emp}, \Gamma_{4, emp}$) (equation (\ref{b1d})), for the GARCH-double-normal model. The intercept of the two lines shows the value of $\alpha_0$ which allows us to model data for a certain median time and a certain time window within the GARCH-double-normal(1,1) model.}
\label{intercept}
\end{figure}

Now for each desired data set we can use the trajectories in the same manner as we have done with the GARCH-normal model. We can plot $\Gamma_6$ along the trajectories of ($\alpha_1$, $\beta_1$) using the running parameter $\alpha_0$, overlaying this with the empirical value we gain figure \ref{intercept}.

\begin{figure}[h!]
\centering
\begin{subfigure}{0.5\textwidth}
\centering
\includegraphics[width=\textwidth]{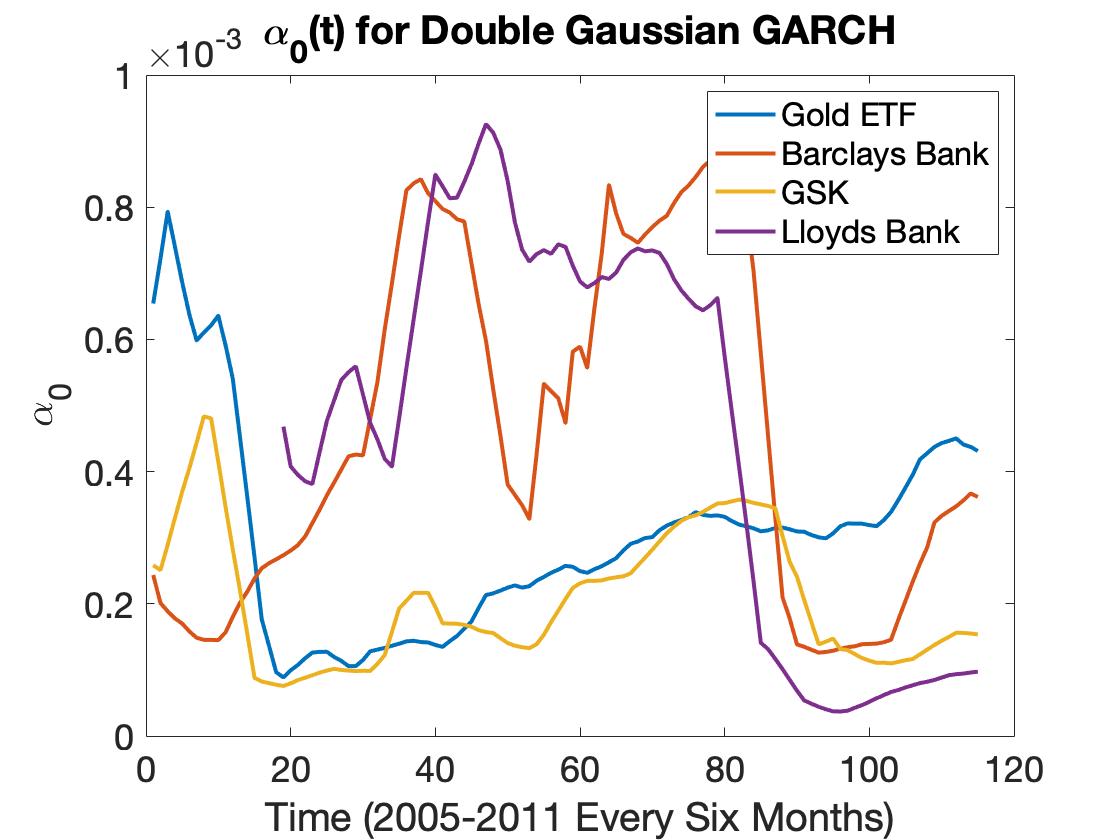}
\caption{}
\label{a}
\end{subfigure}%
\begin{subfigure}{0.5\textwidth}
  \centering
  \includegraphics[width=\linewidth]{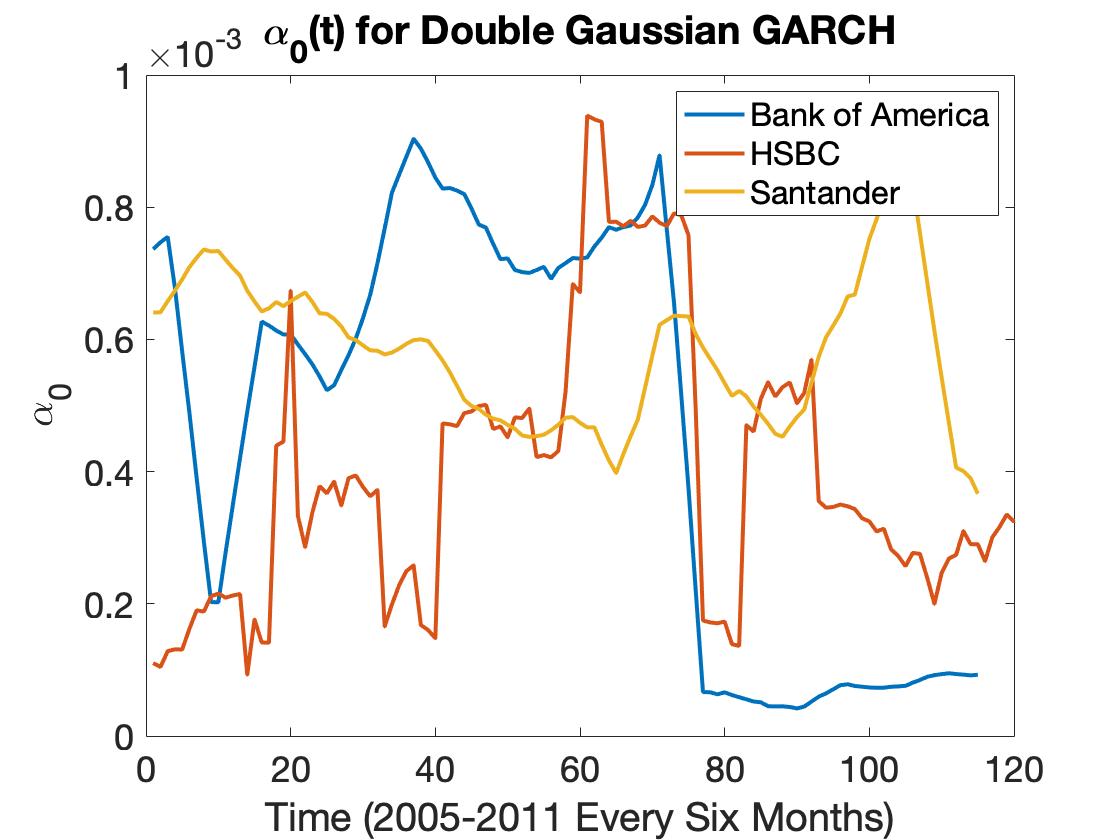}
  \caption{}
  \label{b}
\end{subfigure}
\caption{$\alpha_0(t)$ for several companies between 2005 and 2011. We show here the evolution of $\alpha_0$ when we calculate the parameter value in the shifting 6 month time window. In panel (a), we show the time evolution of $\alpha_0$ for Barclays Bank, Gold ETFs, GSK and Lloyds Bank, by the red, blue, yellow and purple lines respectively. Whilst in panel (b), we show the Bank of America's evolution of $\alpha_0$ by the blue line, the evolution of HSBC by the red and Santander's by the yellow line.}
\label{a0_double}
\end{figure}

From the above method we can recover the value of $\alpha_0(t)$ that allows the fitting of $\sigma^2_{emp}(t)$, $\Gamma_{4, emp}(t)$ and $\Gamma_{6, emp}(t)$, where $t$ is the median of the running window, allowing us to create figure \ref{a0_double}. This is done for several banks: Lloyds Bank, Barclays Bank, Bank of America, HSBC and Santander, a commodity, Gold ETFs and a pharmaceutical company, GSK. We seek to find a fingerprint of the companies' GARCH parameters through the financial crisis. It is evident from figure \ref{a0_double} that the banking sector has a unique behaviour in response to the crisis. We see an initial fluctuating signal, but when the crisis period occurs we see a `cliff-edged' drop in the parameter value. This behaviour is not seen in the commodity, Gold, or the pharmaceutical company stock, GSK, and so we infer that this is due to the response to the crisis period of 2008. It is our belief that this specific behaviour exhibited by the banking companies can be used as an indicator for future banking crisis periods.

\section{Conclusion}
\label{conc}
We use the time series of The Bank of America, Barclays Bank, Citi Bank, HSBC, Gold ETFs, GlaxoSmithKlein and Lloyds Bank, among others, to highlight the inability of the gaussian conditional distribution within a GARCH(1,1) model to fit higher order moments of market time series.

In discovering this, we turn our attention towards different conditional distributions to try to capture the empirical data's higher order moments. We show that with the use of a GARCH-double-normal model we can fit the empirical data's higher order moments. However, through this enquiry, we still cannot capture the long run dynamics of the empirical data. We show that it is only possible to fit a model to empirical data within certain time horizons. To model a different time horizon we have to change the parameters of the double gaussian distribution we use.

Fixing the distribution within certain time horizons to enable the fitting of higher order moments, highlights that the obtained GARCH-double-normal(1,1) model describes a non-stationary process. Therefore, if we wish to describe a long time series by a GARCH-double-normal model, we have to truncate it to smaller time windows. In doing so, we have to potentially fit GARCH-double-normal models with different parameters ($\alpha_0$, $\alpha_1$ and $\beta_1$) to each time window. Therefore, we produce a time dependence of the GARCH model's parameters, for example, $\alpha_0$. As such, we are able to build up a time signature of the $\alpha_0$ parameter through the 2008 financial crash for several companies. We focus our attention on the banking sector to distinguish any shared behaviour in the evolution of $\alpha_0$, through this crisis period. The banking companies' values of $\alpha_0$ have a distinct behaviour from other sectors of the economy, giving hope of a standardised signal of these periods. It is seen through the banking sector's empirical data that before the financial crash there is an increase in $\alpha_0$ and during the financial period, the value of $\alpha_0$ reduces extremely quickly. A behaviour that is found among banking companies but not other securities' time series. This finding is potentially useful for either forecasting or predicting financial turbulence in economic periods.

\section{Acknowledgements}
The authors would like to thank Professor Alistair Milne for his guidance and advice on this work and Laura Stuffins for her preliminary work on the stability of higher order moments in GARCH.

 \bibliographystyle{IEEEtran}
 \bibliography{projectdatabase}

\appendix
\section{Divergence Line Expressions}
\label{anal_divergence}
For the fourth and sixth order moment we can obtain the divergence line explicitly, and so derive:

\begin{equation}
\beta_1^{(4)} = \sqrt{1-2\alpha_1^2} - \alpha_1
\end{equation}

\begin{equation}
\beta_1^{(6)} = \frac{(-8\alpha_1^3+\sqrt{96\alpha_1^6 - 16\alpha_1^3+1}+1)^{\frac{1}{3}}}{2^{\frac{1}{3}}}-\alpha_1-\frac{2(2)^{\frac{1}{3}}\alpha_1^2}{(-8\alpha_1^3+\sqrt{96\alpha_1^6 - 16\alpha_1^3+1}+1)^{\frac{1}{3}}}
\label{cardanos}
\end{equation}
For higher order moments, the divergence lines are defined by high order algebraic equations, which cannot be solved analytically.

\section{Conditions for $\Gamma_4$}
\label{divergence_proof}

For a general GARCH conditional probability distribution with variance equal to one, the equation for the sixth order divergence line (the denominator of equation (\ref{6})) becomes:

\begin{equation}
1-\beta_1^3-3\alpha_1\beta_1^2-3\eta_4\alpha_1^2\beta_1-\eta_6\alpha_1^3 = 0
\end{equation}
Expanding $\beta_1$ in a series with respect to $\alpha_1$ we derive:

\begin{equation}
\beta_1 = 1-A\alpha_1-B\alpha_1^2-C\alpha_1^3-......
\end{equation}
Substituting this into our sixth order divergence line we can equate coefficients up to the second order and so $\beta_1$ becomes:

\begin{equation}
\beta_1 = 1-\alpha_1-(\eta_4+1)\alpha_1^2+O(\alpha_1^3)
\end{equation}
If we now neglect $\alpha_1$ orders higher than the second, we get the equation; $\beta_1 = 1-\alpha_1-(\eta_4+1)\alpha_1^2$. Substituting this into our equation for the fourth order standardised moment, we obtain:

\begin{equation}
\Gamma_4 = \frac{\eta_4(1-\alpha_1-(1-\alpha_1-(\eta_4+1)\alpha_1^2))}{1-\eta_4\alpha_1^2-2\alpha_1(1-\alpha_1-(\eta_4+1)\alpha_1^2)-(1-\alpha_1-(\eta_4+1)\alpha_1^2)^2}
\end{equation}

Considering the limit when $\alpha_1 \to 1$ we finally obtain:

\begin{equation}
\lim_{\alpha_1\to 1} \Gamma_4 = 2\eta_4
\end{equation}

\section{Relations between the parameters of the double gaussian distribution and its higher order moments}
\label{double_gaussian}
The normalisation condition for the double gaussian distribution described by equation (\ref{p_double}) is $a + b = 1$.
Substituting $a=1-b$ into the equation (\ref{double_2}), $E[x^2]$, we get:

\begin{equation}
b = \frac{1-\sigma_1^2}{\sigma_2^2-\sigma_1^2}
\end{equation}
and,

\begin{equation}
a = \frac{\sigma_2^2-1}{\sigma_2^2-\sigma_1^2}
\end{equation}

Assuming $\sigma_1^2 < 1 < \sigma_2^2$, and substituting the equations for $a$ and $b$ into the fourth and sixth moment equations we derive:

\begin{equation}
\mu_4 = \sigma_2^2+\sigma_1^2-\sigma_1^2\sigma_2^2
\end{equation}
and,

\begin{equation}
\mu_6 = (\sigma_1^2+\sigma_2^2)^2-\sigma_1^2\sigma_2^2-\sigma_1^2\sigma_2^2(\sigma_1^2 + \sigma_2^2)
\end{equation}
where, $\mu_4 = \frac{\eta_4}{3}$ and $\mu_6 = \frac{\eta_6}{15}$. Introducing the new variables, $X = \sigma_2^2+\sigma_1^2$ and $Y = \sigma_1^2\sigma_2^2$, we can simplify the obtained equations: 

\begin{equation}
Y = X-\mu_4
\end{equation}

\begin{equation}
X(\mu_4-1)+\mu_4 = \mu_6
\end{equation}
Solving the above equations for $X$ and $Y$ we finally obtain:

\begin{equation}
X = \frac{\mu_6-\mu_4}{\mu_4-1}
\end{equation}

\begin{equation}
Y = \frac{\mu_6-\mu_4^2}{\mu_4-1}
\end{equation}
Since $X$ and $Y$ must be positive, this gives us three conditions; $\mu_4>1$, $\mu_6>\mu_4$ and $\mu_6>\mu_4^2$. Due to the first one we can disregard the second as $\mu_4^2>\mu_4$. Using relations between $\mu_6$ and $\mu_4$ and $\eta_4$ and $\eta_6$ we obtain:

\begin{equation}
\eta_4 >3
\label{con1}
\end{equation}

\begin{equation}
\eta_6 > \frac{15}{9}\eta_4^2
\label{con2}
\end{equation}

We can then set-up equations for solving $\sigma^2_1$ or $\sigma^2_2$:

\begin{equation}
\sigma_2^4 -X\sigma_2^2+Y = 0
\end{equation}
and, 

\begin{equation}
\sigma_1^2 = \frac{Y}{\sigma_2^2}
\end{equation}
Solving for $\sigma_2^2$ we can obtain relations for the parameters of the double gaussian distribution:

\begin{equation}
\sigma_2^2 = \frac{1}{2}(X+\sqrt{X^2-4Y})
\end{equation}
and so,

\begin{equation}
\sigma_1^2 = \frac{2Y}{(X+\sqrt{X^2-4Y})}.
\end{equation}

Since, $\sigma_1^2$ and $\sigma_2^2$ must be both real and positive, this gives us the relation; $X>4Y$. As such we get the following inequality:

\begin{equation}
\mu_6^2-6\mu_4\mu_6-4\mu_6-\mu_4^2(3-4\mu_4)>0
\label{cond}
\end{equation}
Solving this inequality for $\mu_6$, we get the condition; $\mu_4 > -1$. Obviously, $\mu_4$ is always larger than $-1$, and so we always satisfy the condition shown in equation (\ref{cond}.) As such, the parameters $\eta_4$ and $\eta_6$ have to only obey the conditions shown in equations (\ref{con1}) and (\ref{con2}).

\end{document}